\documentclass[10pt]{article}
\usepackage{template}
\begin{document}
\title{\textsf{SpreadNUTS} --- Moderate Dynamic Extension of Paths for No-U-Turn Sampling \& Partitioning Visited Regions}
\date{May 17, 2023}
\author{Fareed Sheriff}
\maketitle
\section{Introduction \& Prior Work}
\par Markov chain Monte Carlo (MCMC) methods have existed for a long time and the field is well-explored. The purpose of MCMC methods is to approximate a distribution through repeated sampling; most MCMC algorithms exhibit asymptotically optimal behavior in that they converge to the true distribution at the limit. However, what differentiates these algorithms are their practical convergence guarantees and efficiency. While a sampler may eventually approximate a distribution well, because it is used in the real world it is necessary that the point at which the sampler yields a good estimate of the distribution is reachable in a reasonable amount of time. Similarly, if it is computationally difficult or intractable to produce good samples from a distribution for use in estimation, then there is no real-world utility afforded by the sampler. Thus, most MCMC methods these days focus on improving efficiency and speeding up convergence.
\par We present a cursory overview of popular MCMC techniques. Random-walk Metropolis-Hastings is a rudimentary algorithm for sampling from a distribution by inducing a Markov chain on repeated samples: the next sample is chosen through a draw from the sampling distribution that takes the current sample as a parameter. However, as the name suggests, this exhibits strong random walk behavior, making it undesirable practically due to the possibly long burn-in period and large number of samples needed to thoroughly explore the distribution space. In fact, many MCMC algorithms suffer from random walk behavior and often only mitigate such behavior as outright erasing random walks is difficult. Hamiltonian Monte Carlo (HMC) is a class of MCMC methods that theoretically exhibit no random walk behavior because of properties related to Hamiltonian dynamics. This paper introduces modifications to a specific HMC algorithm known as the no-U-turn sampler (NUTS) that aims to explore the sample space faster than NUTS, yielding a sampler that has faster convergence to the true distribution than NUTS.
\subsection{Hamiltonian/Hybrid Monte Carlo}\footnote{This subsection summarizes relevant parts of \cite{ham}}
\par Hamiltonian dynamics work on a system of position-momentum pairs $(p,q)$ subject to Hamilton's equations
\begin{equation*}
	\frac{dq_i}{dt} = \frac{\partial H}{\partial p_i},\quad \frac{dp_i}{dt} = -\frac{\partial H}{\partial q_i}
\end{equation*}
where $p,q$ are vector-valued functions of time over a $d$-dimensional space and $H(q,p)$ is the Hamiltonian, which represents the system's total energy. We assume for HMC that the Hamiltonian expresses the system's potential and kinetic energies $H(q,p) = U(q)+K(p)$. We also define for HMC $U(q)$ to be the negative of the log density of $q$ up to a constant and $K(p) = \sfrac{1}{2}p^TM^{-1}p$ to be the negative of the log density of the Gaussian with zero mean and covariance matrix $M$ (often, the Gaussians will be uncorrelated, so $M$ will be diagonal), also up to a constant. We thus rewrite Hamilton's equations to be
\begin{equation*}
	\frac{dq_i}{dt} = (M^{-1}p)_i,\quad \frac{dp_i}{dt} = - \frac{\partial U}{\partial q_i}
\end{equation*}
\par As with MCMC methods as a whole, the Hamiltonian is (time-)reversible and is invariant under Hamilton's equations, meaning the acceptance probability is 1. In practice, it is close to 1 because we cannot practically make the Hamiltonian invariant when solving Hamilton's equations due to error accumulated when solving the PDEs numerically.
\par To numerically solve the PDEs, we use a symplectic integrator, which preserves the Hamiltonian's invariance under integration of Hamilton's equations. A commonly-used symplectic integrator is the leapfrog integrator, which makes use of a "halfstep" in the integration process to better inform the estimate of the Hamiltonian in the next timestep. The equations that govern the leapfrog integrator are as follows with stepsize $\eps$:
\begin{equation*}
	p_i(t+\sfrac{\eps}{2}) = p_i(t)- \frac{\eps}{2}\frac{\partial U}{\partial q_i}q(t)
\end{equation*}
\begin{equation*}
	q_i(t+\eps) = q_i(t) + \eps \frac{p_i(t+\sfrac{\eps}{2})}{m_i}
\end{equation*}
\begin{equation*}
	p_i(t+\eps) = p_i(t+\sfrac{\eps}{2}) - \frac{\eps}{2} \frac{\partial U}{\partial q_i} q(t+\eps)
\end{equation*}
In effect, we compute an estimate of $p$ at $t+\sfrac{\eps}{2}$, estimate $q$ using this estiamte of $p$, then again estimate $p$ using the estimate of $q$ at $t+\eps$, thus taking into account the estimate of $p$ at $t+\sfrac{\eps}{2}$ and $p$'s relationship with $q$.
\par HMC samples from continuous distributions on $\R^d$ with well-defined densities and partials of the log densities. We define the joint distribution $P$ of $(p,q)$ on the Hamiltonian $H$ to be 
\begin{equation*}
	P(q,p) = \frac{1}{Z}e^{-\frac{1}{T}H(q,p)}
\end{equation*}
for any positive constant $Z$ and $T$. Then,
\begin{equation*}
	H(q,p) = U(q)+K(p) \rightarrow P(q,p) = \frac{1}{Z}e^{-\frac{U(q)}{T}}e^{-\frac{K(p)}{T}}
\end{equation*}
We choose $U(q)$ to be $-\log{\pi(q)}$ for the distribution $\pi$ from which we are trying to sample. The distribution of $K(p)$ is independent of $q$, but it is common to use a quadratic like $K(p) = \frac{p^TM^{-1}p}{2}$. For diagonal $M$, this yields $K(p) = \sum_i{\frac{p^2_i}{2m_i}}$.
\par HMC works in two steps. The first step draws a value for momentum $p$ using the zero-centered Gaussian with covariance matrix $M$. The second step conducts a Metropolis update using the Hamiltonian. Using a stepsize of $\eps$ for $L$ steps, a trajectory of samples is calculated, which is accepted with probability
\begin{equation*}
	\min\left(1,\exp\left({\underbrace{U(q)-U(q^*)+K(p)-K(p^*)}_{H(q,p)-H(q^*,p^*)}}\right)\right)
\end{equation*}
which works exactly because the Hamiltonian is time-reversible.
\par Practical considerations to take into account when implementing HMC include varying $\eps,L$. Note, however, that HMC requires adjustment/setting of the parameters $\eps, L$.
\section{No-U-Turn Sampling}%\footnote{This subsection summarizes relevant parts of \cite{nuts}}
\par One of the few and biggest problems with HMC\cite{ham} is the necessity to tune $\eps,L$ --- without proper tuning, we lose many of the efficiency guarantees of HMC. No-U-turn sampling (NUTS)\cite{nuts}] aims to alleviate some of these problems. NUTS is a type of HMC algorithm that does not calculate the trajectory for constant $L$ steps and instead stops the trajectory when sufficient error or space explored has been accumulated. Furthermore, it tunes $\eps$ dynamically to make NUTS an effectively parameterless version of HMC.
\par NUTS replaces a constant $L$ by stopping the trajectory once some condition has been triggered. This condition is checking that the distance between the proposal $q^*$ and the initial $q$ will not continue to increase. We can check this by taking the product of the momentum and the difference between the sampled proposal and initial proposal $(q^*-q)\cdot p^*$ (the U-turn condition), noting that if it is negative, then the direction of our next step will be toward already-sampled points. Because this does not maintain time-reversibility, NUTS runs the Hamiltonian both forward and backward with equal probability and calculates the U-turn condition between the endpoints of the extension of the trajectory generated in the current iteration, checking that it is nonnegative. NUTS generates the trajectory through a doubling scheme that randomly chooses a direction (forward or backward in time), then on the $i$th iteration of generating this trajectory takes $2^i$ timesteps in the chosen direction, adding the calculated points to the current trajectory. A point is chosen as a sample from this trajectory in the following manner: once the trajectory is generated first by sampling some rejection energy threshold $u$ uniformly from $[0,P(q,p)] = [0,e^{-H(q,p)}]$, extending the point forward and backward in time repeatedly, then uniformly randomly selecting a point from this "tree" of points (trajectory).
\section{Moderate Dynamic Extension of Paths}
\par We consider two additions to the NUTS scheme: relaxing the U-turn condition checks on the induced binary tree of the generated trajectory with, and increasing the size of the trajectory by more than double every iteration. Our reasoning behind both of these ideas is that the number of U-turn condition checks on the subtrees of the subtrajectory created by the doubling process in NUTS adds excessive (and potentially underjustified) overhead when checking that the U-turn condition is not violated between the two leaves on the edge of each sutree. This overhead is linear in the number of generated points. While it is stated that "except for very simple models with very little data, the costs of these inner products should be negligible compared to the cost of computing gradients" \cite{nuts} (in reference to the inner products calculated when evaluating the U-turn condition), such a rigorous check can in and of itself be counterproductive and could risk cutting off the trajectory being generated before it has sufficiently explored the space around it. This is because while the U-turn condition checks whether the trajectory turns back on itself, if we check for violation between many pairs of points, adjacent or not, this degenerates into a check that the trajectory is always pointing in the direction of unexplored space.
\par However, this is not a very useful condition to force because we could have a trajectory that moves backward a tiny bit but later continues to move away from previously-explored points, thus exhibiting a general trend toward unexplored space. While we agree that checking that no violation of the U-turn condition should occur between the first few points on the path, we note that as long as the general trend of the path does not violate the U-turn condition, the path contributes to exploring space. We thus strike a compromise: we relax the U-turn condition checks on the balanced tree built on each iteration's points by continuing to check that the U-turn condition is not violated between the leaves on the edge of each subtree of the tree built on each iteration's point, but now build a $k$-ary tree on the calculated points instead of a binary tree where $k$ is the iteration number. This both decreases the number of U-turn condition checks and iteratively relaxes the strictness of the U-turn violation penalty as more points are generated.
\par Specifically, instead of doubling the tree by adding $2^k$ points to the end of our path in direction $d\sim\Uniform{\{-1,1\}}$, we add $k^k$ points and check the U-turn condition fewer times on these points: where we would check the U-turn condition around $2^{k\log_2{k}}$ time on these $k^k$ points, we now check the condition $\frac{k^{k}-1}{k-1}\approx k^{k-1}=2^{(k-1)\log_2{k}}$, which is less than $2^{k\log_2{k}}$ by a multiplicative factor of $k$ (which grows asymptotically).
\section{Partitioning Visited Regions}
\par To prevent ourselves from exploring parts of the distribution that we have already explored, when sampling from the generated trajectory, we bias our selection toward points the space around which we have not already explored. This still satisfies detailed balance because the probability of having already chosen a point from some subspace of the distribution is uniform across all subspaces. Thus, we still have the same convergence guarantees as NUTS. However, we attempt to sample the distribution in a more "spread out" manner by exploring unexplored parts of the trajectory (which itself maintains the invariant of a fixed density) so in the end we still sample in accordance with the distribution's density but with regularization that enforces exploring unexplored parts of the space.
\par We can keep track of how much space we have explored close to a datapoint using any type of querying data structure that allows us to calculate some measure of how explored the space around a given point is (for example, a multidimensional Gaussian convoluted with all previously-sampled points). For sake of example and efficiency, we consider a $k$-dimensional binary search tree $T$ on all sampled points that allows us to find the closest point in average-case $\O(\log{n})$ time with insertion also taking $\O(\log{n})$. %\footnote{This assumes points are randomly distributed, which we assume to be true for now.}.
Our metric $d_p$ for how much space has been explored near a given point $p$ will be the squared $L_2$ norm of $p$ with the closest neighbor in $T$ (sum of squares of difference of coordinates). We then define the probability of choosing $p$ to be proportional to $d_p$ and the metric on all other points of the trajectory so that the probability we select $p$ from trajectory $t = (p_0,\cdots, p_k)$ equals
\begin{equation*}
	\frac{m_p}{\sum_{p_i\in t}{m_{p_i}}}
\end{equation*}
We can then choose a point by allocating some proportion of a uniform r.v. to each point and sampling from this uniform to select the point. This is efficient and so the entire procedure allows us to regularize toward sampling the distribution thoroughly while maintaining sampling by density with the cost of a multiplicative $\O(\log{n})$ factor to the sampling process.
\section{Results}
\par We discuss our testing regime in more detail: we randomly generate mixtures of multivariate Gaussians, which we use to compare how well regular NUTS samples compared to the modified NUTS algorithm presented in this paper by comparing the empirical distributions of each algorithm with the true distribution of the mixtures using a sort of discretized total variation metric. We refer to our algorithm as "SpreadNUTS" because it attempts to spread NUTS trajectories over the sample space to better leave less of the sample space unexplored.\footnote{Our code for SpreadNUTS is based on the code at \cite{impnuts}, and we test SpreadNUTS against this implementation of NUTS}
\subsection{Testing Regime}
\par We randomly select $k$ Gaussian distributions where $k$ is distributed over a discrete uniform that takes values from 1 to 4 (the choice of 5 is arbitrary). We choose the means of the distributions uniformly randomly from the interval $[\vec{-20}, \vec{20}]$ (this choice is also arbitrary); we choose the covariance matrix by generating a matrix whose entries are uniformly random over $[0,1]$, multiplying it by its transpose (generating a valid correlation matrix), then multiplying by a draw from a uniform over interval $[0,4]$ (also arbitrary). This ensures the covariance matrix is positive semidefinite (and is also diagonally dominant). We also uniformly randomly choose a dimension for the Gaussians from 1 to 5. Finally, we generate mixture probabilities $\vec{p}$ such that the elementwise sum is 1 and each value is nonnegative by generating $\Uniform[0,1]$ entries, then dividing by the sum of these entries. While this does not yield a uniform distribution (the distribution is biased toward $\vec{\sfrac{1}{D}}$ where $D$ is the dimension and is chosen uniformly from 1 to 3 --- the low upper bound on dimension is because for dimensions 4 or higher, regular NUTS tends to perform very slowly and it takes too much time to generate samples), this is okay for our purposes because we desire mixtures biased toward uniformly sampling from each vertex so there is sufficient density for sampling algorithms to actually sample from the Gaussians. This randomly generates Gaussian mixtures. Our choice of using Gaussian mixtures was arbitrary and based primarily on convenience of sampling through methods other than Monte Carlo.
\par We define our discretized total variation metric by randomly sampling from the Gaussian mixture (which we do by randomly sampling from each Gaussian, then choosing a number of samples from each collection of samples proportional to the probability of the Gaussian relative to the rest of the mixture). We then generate a relative empirical pdf by discretizing the interval from $\vec{-20}$ to $\vec{20}$ into $0.1$-unit squares, calculating the proportion of samples in each square. Our discretized total variation metric $m_{TV}$ is calculated by taking the absolute difference between the relative empirical pdfs of the samples generated from each algorithm and the relative empirical pdf generated by sampling directly from Gaussians weighted by the relative empirical pdf of the Gaussians. Our comparison between the two algorithms is done by both looking at both the ratio and actual values of $m_{TV}$ between the algorithms and the mixture samples over choice of dimension. We also compare this with the $m_{TV}$ between the Gaussian mixtures resampled again in order to obtain a means of roughly evaluating how well our algorithm performs both relative to NUTS and relative to a true sampler.
\subsection{Results \& Conclusion}
\par We compare the $m_{TV}$ metric between NUTS and SpreadNUTS by plotting them against each other and samples resampled from the mixture as well as by plotting the log of the $m_{TV}$ ratio between NUTS and SpreadNUTS as well as between each algorithm and samples resampled from the mixture. In the first plot, the lower the $m_{TV}$, the better. In the second plot, the close to 0 the score the better; specifically, the log of the ratio between the algorithm and resampled mixture should ideally be close to 0 because this indicates it performs as well as samples from the mixture. We then discuss trends we noticed and provide examples of plots to compare NUTS to SpreadNUTS visually.
\begin{figure}[H]
    \centering
    \includegraphics[width=7cm]{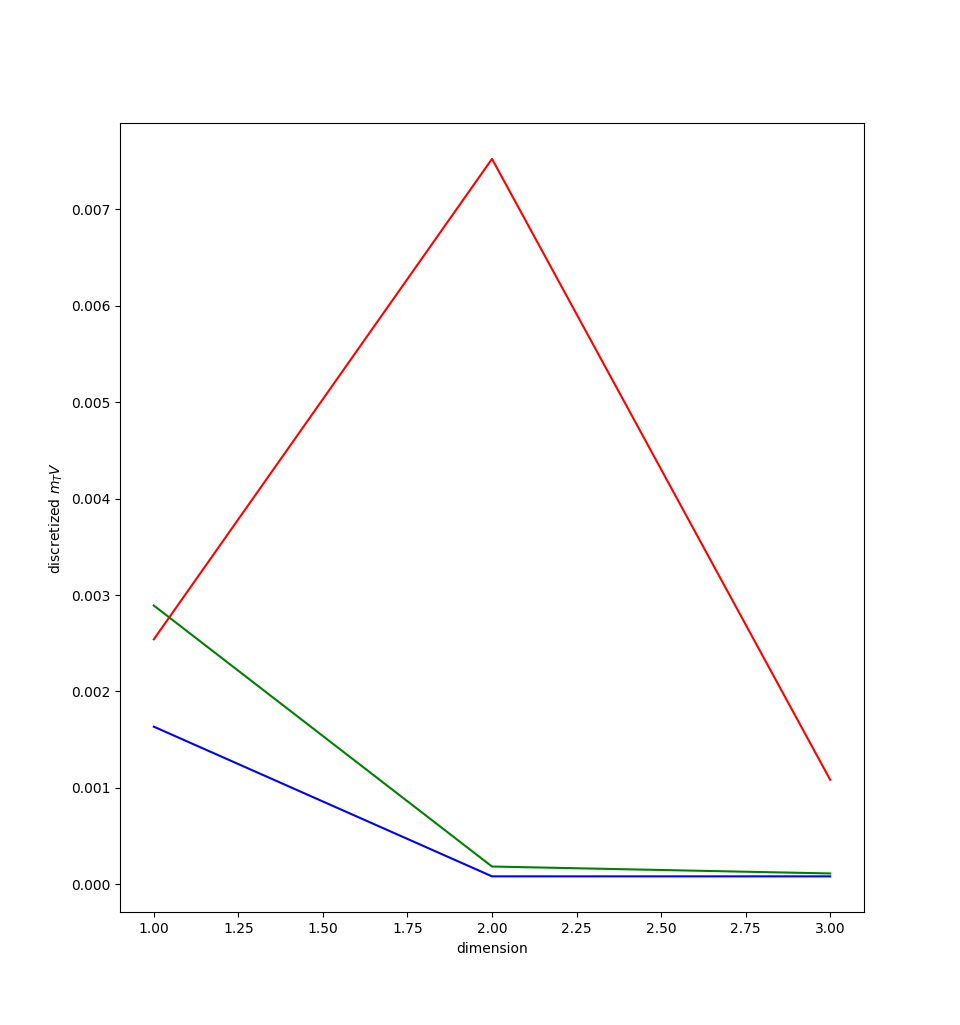}
    \includegraphics[width=7cm]{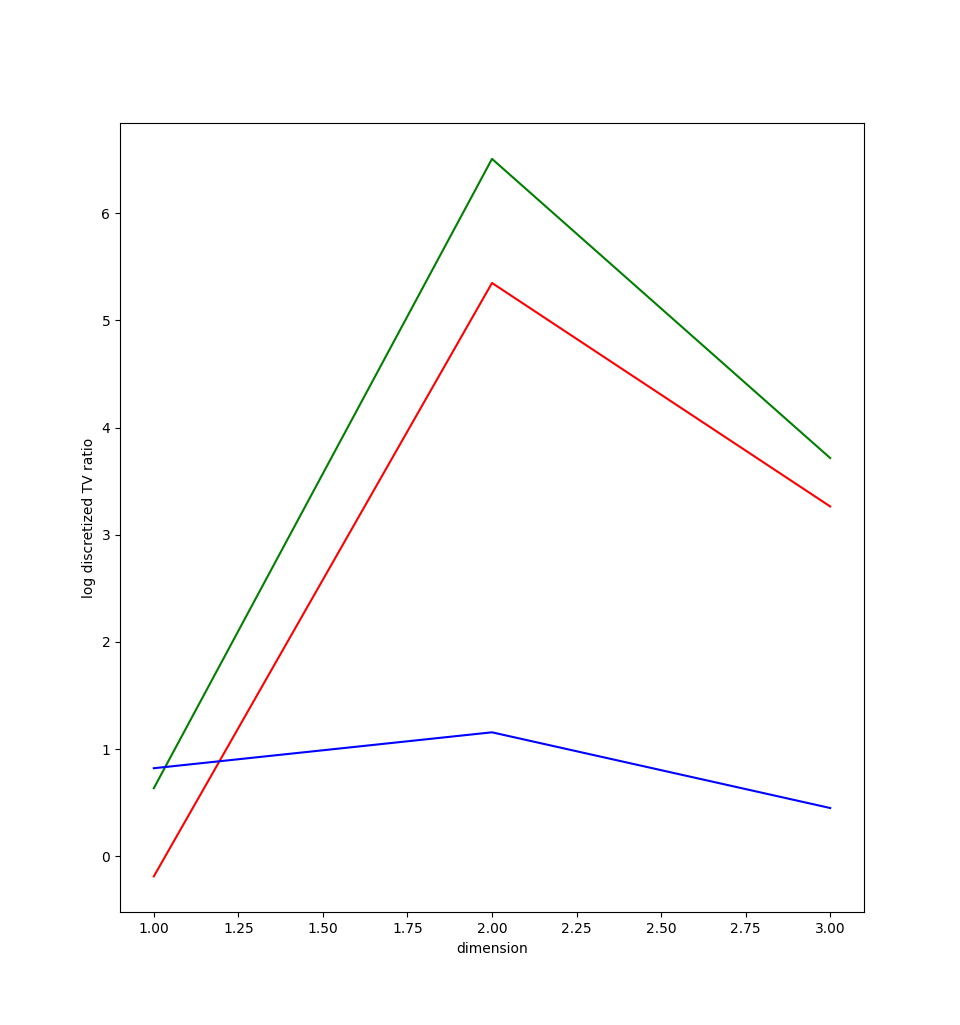}
	\caption{Left: $m_{TV}$ of NUTS (red), SpreadNUTS (green), sampling (blue); right: log of ratio of $m_{TV}$ between NUTS and SpreadNUTS (red), NUTS over sampling (green), SpreadNUTS over sampling (blue)}
\end{figure}
\par The following is a plot of $m_{TV}$ vs. dimension for NUTS, our algorithm, and samples from a Gaussian mixture all compared against samples from a Gaussian mixture. Note that we compare two distinct draws from a Gaussian mixture with each other when calculating the $m_{TV}$ to estimate how much of the $m_{TV}$ of the algorithms is due to randomness attributed to relatively small sample size (we sample 10000 points per mixture and discard the first 500 as burn-in). Alongside it is a comparison of ratios between NUTS $m_{TV}$ and our algorithm's $m_{TV}$ with the mixture $m_{TV}$ vs. dimension to see how close to a random sample the two algorithms get to $m_{TV}$.
\par The following are plots of $m_{TV}$ ratio with the mixture $m_{TV}$ for varying values of $k$ (the number of Gaussians in the mixture) after fixing dimension.
\par The above shows that for dimension 1, NUTS performs better than SpreadNUTS; however, for higher dimensions, SpreadNUTS gets closer and closer to Gaussian sampling, suggesting that it handles density islands better than NUTS.
\par We note some interesting idiosyncracies of SpreadNUTS: in spite of the fact that it tends to perform better than NUTS in higher dimensions, what might actually be going on is that when the distance between "islands" of density in a distribution is sufficiently small enough for classical NUTS to feasibly leap across islands, SpreadNUTS simply makes it more likely that we will actually leap across islands. However, when the distance between these islands is too large for classical NUTS to reasonably travel between islands, SpreadNUTS cannot increase a low probability of traversing these islands enough for it to happen often. Thus, we conclude that while SpreadNUTS may increase the probability of traversing relatively high-density portions of the distribution relative to classical NUTS, it only attempts to "smooth" sampling across parts of the sample space that classical NUTS explores --- it cannot explore parts of the sample space that classical NUTS does not explore. We examine two examples that showcase this trend: a 2d Gaussian mixture consisting of two distributions $\nrm(\mu,I_2),\nrm(-\mu, I_2)$ with equal weight on both. In the first figure, $\mu = \langle2.5,2.5\rangle$; in the second figure $\mu = \langle5,5\rangle$. We compare SpreadNUTS to NUTS and see that while SpreadNUTS increases the probability of traversing these islands relative to classical NUTS, SpreadNUTS does not traverse the islands when classical NUTS does not. Furthermore, looking at the above figures, we can see that on the whole, SpreadNUTS $m_{TV}$ gets closer to Gaussian sampling as dimension increases while NUTS first increases at dimension 2, then decreases at dimension 3 but still with significantly greater $m_{TV}$ than either Gaussian sampling or SpreadNUTS sampling. We note that the number of dimensions used was small (3) and the number of Gaussians in the mixture was from 1 to 4; furthermore, the number of samples was 9.5K for each sampling method. Some error may have been introduced in the relatively small number of samples. A bigger point of contention is that the number of dimensions was too small to make any concrete claims about the efficacy of NUTS vs. SpreadNUTS and the use of Gaussian mixtures as our sample distribution may have introduced some bias that helps SpreadNUTS sample better than NUTS. There is more testing to be done, but we tentatively conclude that SpreadNUTS alleviates to some degree the lack of sample space exploration present in NUTS.
\begin{figure}[H]
    \centering
    \includegraphics[width=4cm]{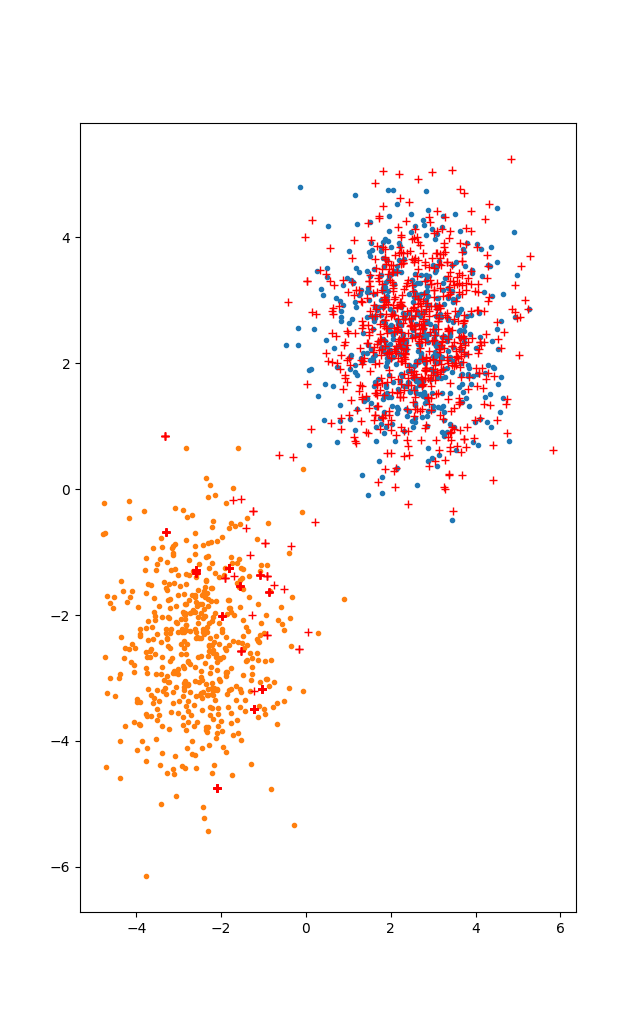}
    \includegraphics[width=4cm]{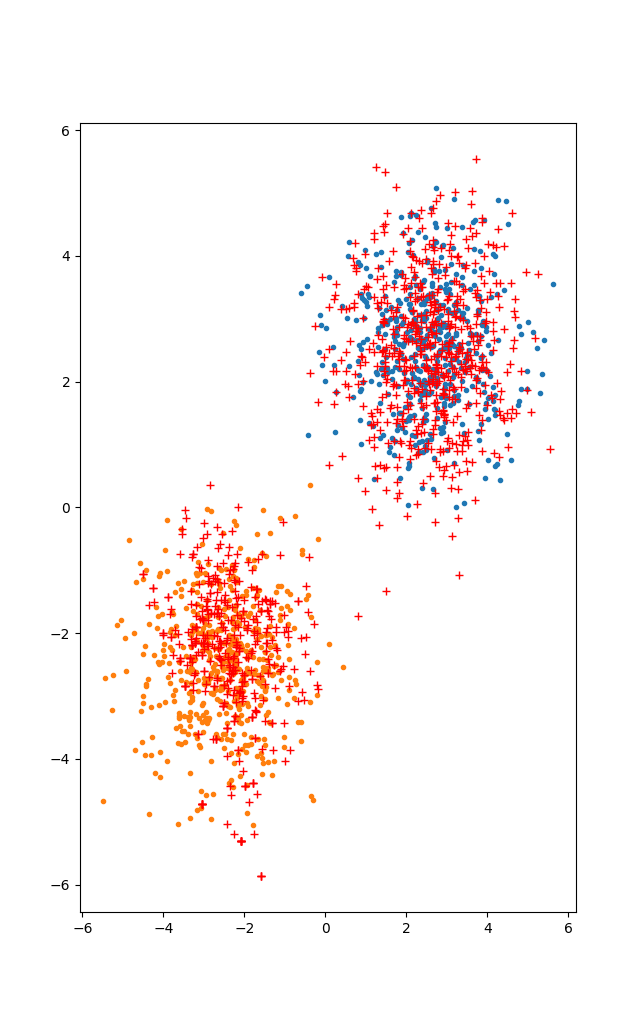}
    \includegraphics[width=4cm]{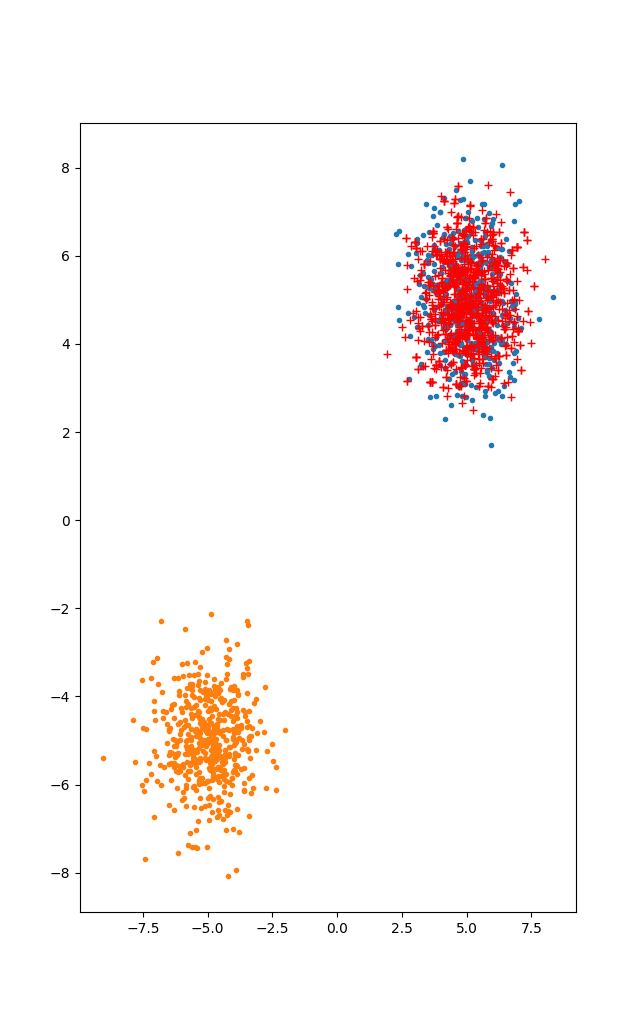}
    \includegraphics[width=4cm]{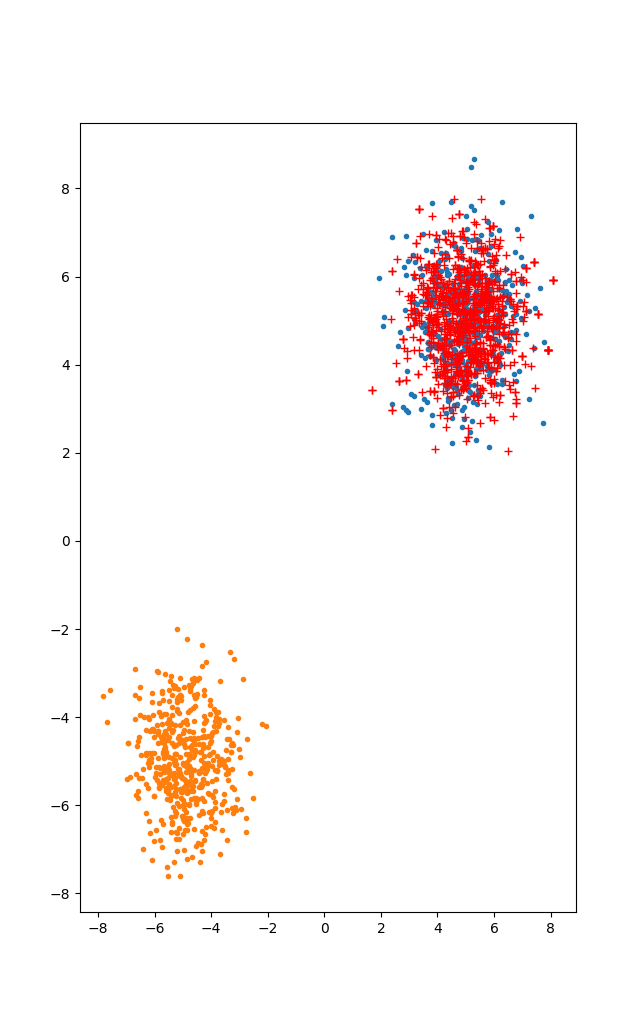}
	\caption{Left is $\mu = \langle2.5,2.5\rangle$, right is $\mu=\langle 5,5\rangle$ with classical NUTS to the left of our algorithm}
\end{figure}
% two means somewhat close to each other, trends include change with dimension & change with k
\newpage
\bibliographystyle{unsrt}
\bibliography{refs}

\begin{thebibliography}{1}

\bibitem{ham}
Brian Keng.
\newblock Hamiltonian monte carlo, 2021.

\bibitem{nuts}
Matthew~D. Hoffman and Andrew Gelman.
\newblock The no-u-turn sampler: Adaptively setting path lengths in hamiltonian
  monte carlo.
\newblock {\em Journal of Machine Learning Research}, 15(47):1593--1623, 2014.

\bibitem{impnuts}
Morgan Fouesneau, Jeremy Sanders, and Muhammad Kasim.
\newblock No-u-turn sampler (nuts) for python, 2020.

\end{thebibliography}
\section{Appendix}
\par We derive the gradient and log-likelihood of Gaussian mixture $M \sim \sum^N{\pi_i\nrm(\mu_i, \Sigma_i)}$. The likelihood (for a single datapoint $x$) is
\begin{equation*}
	p_M(x\mid \pi,\vec{\mu},\vec{\Sigma}) = \sum_{i=1}^N{\pi_i\nrm(x\mid \mu_i,\Sigma_i)}
\end{equation*}
and the log-likelihood is
\begin{equation*}
	\ln{p_M(x\mid \pi, \vec{\mu},\vec{\Sigma})} = \ln\left(\sum_{i=1}^N{\pi_i\nrm(x\mid \mu_i,\Sigma_i)}\right)
\end{equation*}
For a single Gaussian, this devolves to $c -0.5 (\mu- x)^T\Sigma^{-1}(\mu-x)$ for extra constant $c = -0.5\ln(|\Sigma^{-1}|(2\pi)^k)$.
Then, the gradient of the log-likelihood w.r.t. $\vec{\mu}$ is
\begin{equation*}
	\frac{\partial \ln(p_M(x\mid \pi, \vec{\mu}, \vec{\Sigma}))}{\partial \vec{\mu}} = \frac{1}{{\sum_i{\pi_i\nrm(x\mid \mu_i,\Sigma_i)}}} \cdot \frac{\partial p(x\mid \pi,\vec{\mu},\vec{\Sigma})}{\partial \vec{\mu}}
\end{equation*}
\begin{equation*}
	\frac{\partial p(x\mid \pi,\vec{\mu},\vec{\Sigma})}{\partial \vec{\mu}} = \sum_i{\frac{\partial \pi_i\nrm(x\mid \mu_i,\Sigma_i)}{\partial \mu_i}}
\end{equation*}
\begin{equation*}
	\frac{\partial \pi_i\nrm(x\mid \mu_i,\Sigma_i)}{\partial \mu_i} = \frac{\partial}{\partial \mu_i}\left(\pi_i\sqrt{|\Sigma^{-1}_i|(2\pi)^{-k}}\exp(-\frac{1}{2}(\mu_i-x)^T\Sigma^{-1}_i(\mu_i-x))\right) = \Sigma^{-1}(x-\mu_i)\pi_i\nrm(x\mid \mu_i,\Sigma_i)
\end{equation*}
\begin{equation*}
	\frac{\partial \ln(p_M(x\mid \pi, \vec{\mu}, \vec{\Sigma}))}{\partial \vec{\mu}} = \frac{\sum_i{\Sigma^{-1}(x-\mu_i)\pi_i\nrm(x\mid \mu_i,\Sigma_i)}}{\sum_i{\pi_i\nrm(x\mid \mu_i,\Sigma_i)}}
\end{equation*}
\par For a single Gaussian, this simplifies to $\Sigma^{-1}(x-\mu)$.
\par As an aside, our testing regime experiences compounding rounding errors when exponentiating and taking logs, specifically when we take the log of the exponential of a number close to 0, which rounds to 0. We attempt to alleviate this problem by expressing the proportions of the normal likelihoods $\pi_i\nrm(x\mid \mu_i,\Sigma_i)$ to the sum of the normal likelihoods as the exponential of the difference of the log likelihood and the log of the sum of likelihoods, where we calculate the log of the sum of likelihoods by summing logs as below:
\begin{equation*}
	\log(x+y) = \log(x(1+\sfrac{y}{x})) = \log{x} + \log{(1+\sfrac{y}{x})} = \log{x} + \log{(1+e^{\log{y}-\log{x}})}
\end{equation*}
\begin{equation*}
	\log{\sum_i{x_i}} = \log\left(x_1\left(1+\frac{1}{x_1}\sum_{i=2}^k{x_i}\right)\right) = \log{x_1} + \log{(1+e^{\log{\sum_{i>1}{x_i}}-\log{x_1}})}
\end{equation*}
\begin{equation*}
	\log{\sum_{i>1}{x_i}} = \log{x_2} + \log{(1+e^{\log{\sum_{i>2}{x_i}}-\log{x_2}})}
\end{equation*}
\begin{equation*}
	\frac{{x_i}}{{\sum{x_i}}} = \exp(\log{x_i}-\log{\sum{x_i}})
\end{equation*}
Thus, we can recursively express the log of sums as the sum of log sums (in practice, we sort the Gaussian pdfs when evaluating logs to minimize error at each step, yielding a technique known as LogSumExp or LSE). This helps decrease error accumulated when summing likelihoods because of the error introduced when summing exponentials.
\end{document}